\documentclass[12pt,preprint]{aastex}
\usepackage{epsfig, array, amsmath, delarray}

\begin{document}



\title{The 550 AU Mission: A Critical Discussion}


\author{Slava G. Turyshev 
}
\affil{Jet Propulsion Laboratory, California Institute of Technology,\\
Pasadena, CA 91109}

\and

\author{B-G Andersson 
}
\affil{Department of Physics and Astronomy,
The Johns Hopkins University, \\Baltimore, MD 21218}




\begin{abstract}

We have studied the science rationale, goals and requirements for a
mission aimed at using the gravitational lensing from the Sun as a way of
achieving high angular resolution and high signal amplification.  We find
that such a mission concept is plagued by several practical problems. 
Most severe are the effects due to the plasma in the solar atmosphere which
cause refraction and scattering of the propagating rays. These effects
limit the frequencies that can be observed to those above $\sim$100-200 GHz
and moves the optical point outwards beyond the vacuum value of $\geq$ 550 AU.
Density fluctuations in the inner solar atmosphere will further cause
random pathlength differences for different rays.  The corrections for the
radiation from the Sun itself will also be a major challenge at any
wavelength used.  Given reasonable constraints on the spacecraft
(particularly in terms of size and propulsion) source selection as well
as severe navigational constraints further add to the difficulties for a
potential mission.
 
\end{abstract}




\section{Background}

General relativity (GR) predicts that light will be deflected by any massive
object; an effect first experimentally confirmed by \citet{edd19}.   As a
consequence, a far away object will act as a lens because the deflected rays from
the two sides of the lensing mass converge.  This is a well known effect and has
been observed over cosmological distances \citep{bla92} where relatively nearby
galaxies, or even clusters of galaxies, act as gravitational lenses for background
galaxies, and in our Galaxy where micro-lensing of stars in the Galactic bulge or
in the Magellanic clouds are caused by intervening (sub-)stellar bodies
\citep{pac96}.  Even though these studies can yield a wide variety of important
information in many branches of astronomy, we are left at the mercy of Nature in the
selection of sources.  This drawback could in principle be overcome if we could
move a spacecraft to the location of the focus of a gravitational lens due to some
nearby object.  As can be seen from below, of the solar system bodies, only the Sun
is massive enough that the focus of its gravitational deflection is within range of
a realistic mission.\footnote{Grazing incidence rays passed the Sun, if propagation
through vacuum is assumed, focus at a distance of $\sim$550 AU, whereas the optical
length of Jupiter is about 5900 AU.}  The gravitational lens effect of a spherical 
star  for astronomical observations of distant objects has been studied by a number
of authors \citep{ref65,bli75,her76}.  The specific use of the Sun for that purpose
was first proposed in 1979 by von Eshleman, whose original `study' was  aimed at
interstellar communication with a  single  pre-determined pointing at long
wavelengths.  Subsequent discussions include those of \citet{kra86} and
\citet{hei94}.  The main attraction of this technique is the high amplifications
(gain, in astronomy nomenclature) which can be achieved (see \citet{kra86}). 
However, in these studies, several effects of importance to a possible
implementation of such observations were neglected, mainly those due to the solar
plasma and the brightness of the Sun. Herein, we give a brief review of the subject
and present a preliminary investigation of some of the previously ignored effects. 
We also discuss possible observing strategies and source selection.

As with any space mission aimed at reaching distances outside of the solar system
there are a large number of \textit{in situ} measurements which can be performed
during cruise.  Given a suitable suite of instruments, studies of the heliosphere
and its termination as well as the local interstellar medium can be performed.  We
briefly summarize some of the possible areas of cruise science addressed in the
literature.  Further possible secondary science objectives, such as VLBI and
parallax measurements, will not be addressed.

The  paper organized as follows:  In Section \ref{sec:gr} we discuss the basic
parameters of the solar gravitational lens: the optical distance, the amplification
and the gain. We also address the issue of solar plasma contribution to the total
effect.   In Section \ref{sec:sgt}  we discuss the observational considerations for
a concept of solar gravitational telescope. We present the "pros" and "cons" for a
possible mission designed to  utilize the solar gravitational lens for any
scientific application. In Section \ref{sec:cruise} we   present possible
scientific program for the mission during its cruise phase. We conclude with
Section \ref{sec:conclusions}, where we present our conclusions and recommendations
related to the possibility of utilizing the solar gravitational lens in the
future.  

\section{The Sun as a gravitational lens}
\label{sec:gr}
\subsection{GR lensing by a massive object in vacuum}




It is well-known that light rays passing by a gravitating body are deflected by the
body's gravitational field (see \citep{pac96} and references therein). The largest
contribution to the total deflection angle comes from the post-Newtonian monopole
component of the gravity field.   This bending effect (towards the body) depend  on
the mass of the body ${\cal M}$  and the light's impact parameter $b$  relative to
the deflector. For the Sun this effect may be expressed  as:
\begin{equation}
\theta_{\tt gr}(b)= \theta_{{\tt gr}_0}~{{\cal R}_\odot\over b},
\qquad {\rm where} \qquad \theta_{{\tt gr}_0}=-
{2 r_g \over {\cal R}_\odot}=-8.52\times 10^{-6}~\mu{\rm rad},
\label{eqdef}
\end{equation}

\noindent  where $r_g=2G{\cal M}_\odot/c^2=2.95$ km is the Schwarzschild radius of
the Sun. The largest post-post-Newtonian  ($\propto r^2_g/{\cal R}^2_\odot$)
contribution to this first-order bending effect in the solar system is by the Sun
and it amounts to  53 picoradians at the solar limb. Here we shall deal with  one
important part of the wave-theoretical treatment of the gravitational lens, namely,
the scattering of a plane electromagnetic wave by a spherically symmetric star. In
the language of geometric optics, we shall consider the image of a very distant
object. For this approximation the Schwarzschild radius of the gravitational lens
is assumed to be small compared to its radius but large compared to the (flat
space) wavelength of the incident wave.

The gravitational field acts as a lens. However, since for rays passing through the
exterior gravitational fields of the Sun, the deflection angle decreases with
increasing impact parameter (as shown by Eq.(\ref{eqdef})), the lens does not have
a true optical point but only a caustic line beginning at the distance of ${\cal
F}_0 \equiv {\cal F}({\cal R}_\odot) = {\cal R}^2_\odot /2 r_g = 546 ~AU$ from the
Sun. Geometric optics gives the optical distance as a function of the impact
parameter:
\begin{equation}
{\cal F}(b)={\cal F}_0\frac{b^2}{{\cal R}^2_\odot}
\end{equation}
By rigorously applying the methods of wave optics it was shown  by
\citet{her76} that the space behind the Sun may formally be separated into the
three physically different regions, namely, (1) the shadow, (2) the region of
geometric optics (where only one ray passes through each point of space), and (3)
the region of interference (where two rays are passing through each point) as shown
in Figure
\ref{fig:sl}.

The solar shadowing effect prohibits focusing of light  at distances shorter then
${\cal F}_0$ from the Sun. On the other hand, the most interesting effects, such as
amplification of light, may only be observed in the third region -- the region  
interference.  Hence, in discussing the solar gravity lens, we shall be interested
only in the solar region of interference. This region is defined to be at the
distance ${\cal F}$ such as: ${\cal F} \ge {\cal F}_0 $ and $\theta
\le |\theta_{0\,\tt gr}|$. This region may further be sub-divided onto several
physically interesting regions. The most intriguing of those, is the region of
extreme intensity, for which the following condition in the image plane
(perpendicular to the optical axis) is satisfied\footnote{We use the polar
coordinate system ($\rho,\theta$) see Figure \ref{fig:sl}.}:
${\lambda\over 2\pi}/{ \sqrt{2r_g{\cal F}}}
\ll \theta \ll  \sqrt{{r_g/ {\cal F}}}.$

For small departures $\rho$ of the observer from the optical axis
$\rho\ll\sqrt{2r_g{\cal F}}$  solution may be obtained by the stationary phase
method (see \cite{her76}) which yields the following expression for the gain
$G(\rho,\lambda)$ of this lens:
\begin{equation}
G(\rho,\lambda) \cong 4\pi^2 {r_g\over \lambda}J_0^2\left(
{2\pi\rho\over\lambda}\sqrt{{2r_g\over{\cal F}}}\right),
\end{equation}
\noindent $J_0$ being a Bessel function of zero-order.
The corresponding gain as a function of the optical distance and  a possible
observation wavelength is presented in Figure \ref{fig:gain}. Note that gain
$G(\rho,\lambda)$ has it's maximum on the axis
\begin{equation}
G_{\tt max}(0,\lambda)= 4\pi^2 {r_g\over \lambda}
\end{equation}
\noindent for all points on the optical axis, the gain then decreases slowly
(while oscillating) if $\rho$ increases (off-optical line), becoming
\begin{equation}
G(\rho_1,\lambda)\cong \Big({8\pi r_g\over \lambda}\Big)^\frac{1}{2}  \qquad
{\rm when} \qquad \rho_1= \Big({\lambda {\cal F}\over \pi}\Big)^\frac{1}{2}
\end{equation}

\noindent   and going further down to its mean value  $G=1$ for
$\rho \gg \sqrt{{\cal F} \,r_g}$. Thus, as  expected, the gravity lens
is very sensitive to a tangential  motions (e.q. in the image plane). 

Another important feature of the solar gravity lens  is a high angular 
resolution. To describe the angular resolution we will use the
angle, $\epsilon_{10}$, that corresponds to the distance from the optical axis to
the point where gain decreases by 10 dB:
\begin{equation}
\epsilon_{10dB}(b, \lambda)=\frac{\rho_{10dB}}{{\cal F}}\cong
\frac{\lambda}{4\pi\sqrt{2r_g{\cal F}(b)}}=
\frac{\lambda}{4\pi\sqrt{2r_g{\cal F}_0}}\Big(\frac{{\cal
R}_\odot}{b}\Big)
\end{equation}
\noindent where ${\rho_{10dB}}$ is the distance in the image plane 
where the gain decreases by 10 dB (the distance is given by
$\rho_{10dB}\cong\frac{\lambda}{4\pi}\sqrt{\frac{{\cal F}}{2r_g}}$). For  
wavelength $\lambda\sim$ 1 mm  resolution $\epsilon_{10}$ is estimated to
be $\epsilon_{10dB}(b, 1 ~{\rm mm})=0.11~{\cal R}_\odot/{b} 
~\pi {\rm rad} = 0.11~\sqrt{{\cal F}_0/{\cal F}}~\pi$rad.

Since for the Sun the frequency can have values between $10^4$ Hz (radio
waves) and $10^{10}$ Hz (visible light) or even $10^{14}$ Hz 
($\gamma$-rays), a gravitational lens can enlarge the brightness of a star
by the same remarkable factor. The plane wave is focused into a narrow beam
of extreme intensity; the radius of which is 
$\rho=\sqrt{\lambda {\cal F}/\pi}$.  The components of the
Poynting vector and the intensity are oscillating with a spatial period
\begin{equation}
\delta \rho  =  \lambda \Big({{\cal F}\over 8r_g}\Big)^{1\over2} =
1.18 \times 10^5    \Big({\lambda b\over {\cal R}_\odot }\Big).
\end{equation}

\noindent Thus for $\lambda ~\sim$  1 mm  this period is $\delta
\rho =0.118\times b/{\cal R}_\odot $ km.

It is also useful to discuss  the angular distribution of intensity.
Thus, if  telescope is small, the  observed direction to the source will
by determined by the corresponding deflection angle and the
telescope's position. However, if it is large, it will give
a distribution of intensity  ${\cal I}$ over the aperture
\begin{equation}
{d {\cal I}\over d \theta} = \Big({{\cal F}\over
2\rho}\Big)^\frac{3}{2}   {\lambda\rho ~d \theta d \phi\over
\pi\sqrt{\theta}}
\end{equation}

\noindent and the observer would essentially see a significantly magnified
source  at $\theta\cong0$.

\subsection{The effects of the solar atmosphere}

The solar atmosphere introduces several complications to the
above gravitational deflection scenario.  They are due to the facts that it
consists partly of a free-electron gas and that it is turbulent

\subsubsection{Plasma frequency}

First, a free electron gas responds to the (time) variable electric field
of a passing electromagnetic wave and, for low enough frequencies, will
absorb it. The plasma frequency, at which the refractive index of the
plasma turns imaginary and hence absorptive, is proportional to the
square-root of the electron density (e.g. \citet{rub79}) and therefore as we go
progressively lower into the solar atmosphere, progressively higher frequencies
of radio waves will not be able to propagate. \citet{and97} have shown that the
electron density at the Solar photosphere is of the order $(1-5)\times 10^8$
cm$^{-3}$.  So, even for a quiet Sun, no radiation below a few hundred MHz can
propagate through the solar atmosphere.

\subsubsection{Refraction}

Second, the density gradient in the solar atmosphere and hence the
electron density influences the propagation of radio waves through the
medium.  Even for those frequencies which can propagate through it, the
anisotropy of the solar atmosphere will cause refractive bending of the
propagating rays.  Since the density decreases outwards, this will cause a
divergence in the propagating rays.  Hence, the location of the focus for a
given impact parameter will be shifted to larger distances than expected if
the Sun did not have an atmosphere.

Any experiment involving propagation of an electromagnetic waves near the Sun
faces a great challenge to overcome the refraction in the immediate solar
vicinity due to various sources. The solar plasma is the main source of
noise in an observations near the  limb of the Sun. In general, one can
express the total deflection angle $\theta_{\tt tot}$ as follows:
\begin{equation}
\theta_{\tt tot}          = \theta_{\tt gr}+\theta_{\tt pl}+\theta_{\tt n},
\label{eqtotal}
\end{equation}
\noindent where $\theta_{\tt pl}$ is the dominant plasma
contribution, and $\theta_{\tt n}$ contains all non-dispersive sources of 
noise\footnote{We use word 'noise' here to describe the contribution due
to plasma and other sources of measurement errors because, for the purposes of
observing the effect of gravitational deflection, these sources represent the
sources of non-gravitational noise.}
(pointing errors, attitude control, receiver, etc.), along with a minor
dispersive contribution from the interstellar media, asteroid belt, and Kuiper
belt, etc.

The plasma contribution $\theta_{\tt pl}$ to the total deflection angle is
related to the change in the optical path
\begin{equation}
\Delta \ell = {N_e(\ell) e^2\over 2\pi m_e \nu^2},
\end{equation}

\noindent where  $e$ is the electron's charge, $m_e$ it's mass, and
$N_e(\ell)$ is the total columnar electron content along the beam,
$N_e=\int n_e d\ell$. Therefore, in order to calibrate the plasma term, we
should know the electron density along the path. We start by decomposing
the electron density $n_e$ in static, spherically symmetric part
$\overline{n}_e(r)$ plus a fluctuation  $\delta n_e$, i.e.
\begin{equation}
n_e(t, {\bf r})= \overline{n}(r)+ \delta n_e(t, {\bf r}).
\label{eqelcont}
\end{equation}

The steady-state behavior is reasonably well known, and we can use one of
the several plasma models found in the literature
\citep{tyl77,muh77,muh81}. To be more explicit, we will refer to one
particular model, namely (we do not consider here a correction factor due to
the heliographic latitude):
\begin{equation}
\overline{n}_e(r)= \Big[\Big({2.99 \over \eta^{16}} +{1.55 \over
\eta^6}\Big)\times10^8 + {3.44\times 10^5\over \eta^2}\Big],  ~~   {\rm
cm}^{-3} 
\label{eqmodel}
\end{equation}

\noindent where   $\eta=r/{\cal R}_\odot $. At large distances this model
gives the expected behavior $\propto 1/r^2$ of the solar wind.

We will now determine the contribution to the total deflection  angle due to
the solar plasma for the model  above. Following the usual method
outlined in \cite{and97}, we obtain the  corresponding
deflection due to solar plasma $\theta_{\tt pl}$   as follows:
\begin{equation}
\theta_{\tt pl}(b,\nu) =\left({\nu_0  \over \nu}\right)^2 \left[ ~2.952
\times  10^3
\Big({{\cal R}_\odot \over b}\Big)^{16}  +  2.28 \times  10^2 \Big({{\cal
R}_\odot \over
b}\Big)^6  +  1.1  \Big({{\cal R}_\odot \over b}\Big)^2 \right]
\end{equation}
\noindent with $\nu_0=6.32  ~{\rm MHz}$. Comparing $\theta_{\tt pl}$ with
$\theta_{\tt gr}$  we notice the opposite sign - gravity bends the ray
outwards, plasma inwards -- and the different dependence on $b$, plasma
being steeper. The plasma deflection as a function of the solar offset  $b$
is shown in the Figure \ref{fig:plasma}.

The presence of the solar atmosphere will result also in the de-focusing of the
coherent radiation, thus worsening the quality of the  solar gravitational
lens. To analyze this influence let us   estimate the  optical distance for
the two sources affecting the light propagation in the solar vicinity,
namely gravity and plasma. One may expect that the beginning of the
interference zone will be shifted further away  from the Sun. For
estimation purposes we will consider here only the steady-state part of the
plasma model. The beginning of the interference zone (e.q. the effective
optical distance) in this case may be determined from
\begin{equation}
\frac{b}{{\cal F}_{\tt gr+pl}}=\theta_{\tt tot}=  \frac{2r_g}{b}
-\theta_{\tt pl}.
\end{equation}

\noindent This expression gives the effective optical distance for the system
gravity+plasma ${\cal F}_{\tt gr+pl}\ge 0$:
\begin{equation}
{\cal F}_{\tt gr+pl}(b,\nu)= 546\,\Big(\frac{b}
{{\cal R}_\odot}\Big)^2
\Big[1- \frac{{\cal R}_\odot}{2r_g}\Big(\frac{b}{{\cal
R}_\odot}\Big)\theta_{\tt
pl}(b,\nu)\Big]^{-1}~~{\rm AU}
\label{eqtot}
\end{equation}

\noindent 
One may note that for any given impact parameter $b$ there is a
critical frequency  $\nu_{\tt crit}$ such that the denominator in the
expression (\ref{eqtot}) vanishes and the effective optical distance
${\cal F}_{\tt gr+pl}$ becomes infinitive. This is the case when there is
no lensing at all and the  solar plasma  entirely neutralizes influence of
the solar gravity. This critical frequency is given as follows:
\begin{equation}
\nu^2_{\tt crit}(b) = \Big(2.161~{\rm GHz}\Big)^2 \left[ 2.952 \times  10^3
\Big({{\cal R}_\odot \over b}\Big)^{15}  +  2.28 \times  10^2\Big({{\cal
R}_\odot \over
b}\Big)^5 +   1.1 \Big({{\cal R}_\odot \over b}\Big)\right],
\label{eq:nu}
\end{equation}


Based on the estimates for ${\cal F}_{\tt gr}$ presented in the Table
\ref{tab66} and from the practical considerations for
the solar gravity lens mission,  one will have to limit the  range of possible 
impact parameters to those in the interval
$ {b}/{{\cal R}_\odot } \in [1.05, 1.35]$, which correspond to the optical
distance ${\cal F}_{\tt gr+pl} \in [601, 1000[$ AU. For this range of impact
parameter, the main contribution comes from the term   $\sim
\big({{\cal R}_\odot / b}\big)^{15}$.   Therefore, approximating to the
sufficient  order, we will have the critical frequency
\begin{equation}
\nu_{\tt crit}(b) = \nu_{0\,\tt crit}  ~\Big({{\cal R}_\odot
\over b}\Big)^\frac{15}{2}, \qquad  {b}/{{\cal R}_\odot} \in \Big[1.05,
~1.35\Big]
\end{equation}

\noindent  with 
$\nu_{0\,\tt crit}\equiv\nu_{\tt crit}({\cal R}_\odot )=120~{\rm GHz}$ or,
equivalently,  2.5  mm. As a result of this analysis we find that
the effective optical distance for the system gravity+plasma will be
determined from the following expression:
\begin{equation}
{\cal F}_{\tt gr+pl}(b,\nu)= 546 \,\Big(\frac{b} {{\cal R}_\odot}\Big)^2
\Big[1-  \frac{\nu^2_{0\,\tt crit}}{\nu^2}
\Big({{\cal R}_\odot \over
b}\Big)^{15}  \Big]^{-1} ~~{\rm AU}, \qquad  {b}/{{\cal R}_\odot} \in
\Big[1.05, ~1.35\Big]
\end{equation}

\noindent Results corresponding to a different frequencies are plotted in
Figure \ref{fig:pf}.  This represents  the expected  
solar plasma  response in this region of impact parameter 
$b\in[1.0-1.4]{{\cal R}_\odot}$.
Note that only for the case when  $\nu \gg \nu_{\tt crit}=120$
GHz is it possible to minimize the influence of the solar plasma and to
allow for the geometric optics approximation to estimate the effects of
solar microlensing (with some additional assumptions, for details see 
\citet{her76}). Hence,  from this perspective the observational  wavelength
$\lambda$ for the chosen range of the impact parameters should be much smaller
then
$\lambda_{\tt crit}$.

It is worth  noting that a  solar plasma model  similar to that given by 
Eq.(\ref{eqmodel}) will be used for the Cassini relativity experiments at
the solar conjunction. These experiments will  utilize the dual  frequency
plasma cancellation technique with the telecommunication links available.
For the 550 AU Mission the model for the solar plasma should be known to a
highest degree accuracy including the latitude dependencies in  the plasma
distribution.

\subsubsection{ Turbulence}


In the previous paragraph we have considered a spherical and static Solar
atmosphere model, and derived the resulting plasma contribution to the total
deflection angle. Unfortunately, the true electron density
Eq. (\ref{eqtotal}) contains also the fluctuations   $\delta n_e$, which
require particular attention. In fact, these fluctuations are carried along
with the solar wind speed $V\approx 400$ km/s, so that their scale is  
$V\,\tau  \approx {\cal R}_\odot /2$, where $\tau \approx 10^3$ s  is the
temporal scale. The solar atmosphere is highly variable over all time
scales, due to solar activity; solar flares, coronal holes, solar rotation,
etc. Moreover, on the typical scale of the gravitational deflection of
light, one expects   $\delta n_e$ to be of the same order of magnitude as
its average \citep{arm79}, i.e.  $\delta n_e (t,b)
\approx \langle n_e\rangle(b)$.  In a conservative vein, it seems 
reasonable to assume that the deflection for an impact parameter $b$ is of the same
order as the deflection for the mean solar atmosphere. 

Two further physical optics effects come into play namely: spectral
broadening and angular broadening, however, discussion of these effects is
out of scope of present paper.

\subsection{The Sun's radiation}

Because the Sun is a bright body and the observation of an object,
gravitationally lensed by the sun necessitates pointings at, or close to
the Sun itself, it will dominate the detector system at most
wavelengths.  Therefore, care has to be taken in selecting the optimum
observing frequency and observing strategy.  The integrated solar spectrum
is well approximated by a 6000K black body for wavelengths below about
1cm.  Longward of about 1cm the spectrum deviates from a blackbody and
becomes dependent on the solar cycle. For the active Sun it rises to a
secondary peak at around 1m.  Therefore, a relative minimum exists just
shortward of 1cm. However, even at 1mm the diffraction limited size of an
antenna (of realistic size) will be large compared to the size of the solar
disk.  At 550 AU the sun subtends about 3.5'' while a 10m antenna has a
beam size of (FWFZ) about 50''.  Hence, unless the source to be observed is
strong, observations in the mm-wave range will have to be
differential in wavelength space and have to require that the source has a
different spectrum than the Sun.   This observing strategy imposes
 severe requirements on the stability of the detector system, such that a
reliable subtraction of the solar flux can be achieved.  Coronografic
observations can be envisioned at short (IR, visual) wavelength
observations, however, in the IR and optical, emission and/or scattering
from the Zodiacal dust will present a challenge for such observations. 
Further study is required to quantify these effects. 

\section{Observational considerations, Solar Gravitational Telescope}
\label{sec:sgt}

As noted above, in order to guarantee propagation of the radiation through
the solar atmosphere, observations have to be restricted to frequencies
above the plasma frequency at a given impact parameter as well as above
those where the refraction cancels the gravitational convergence.  If we
want to minimize the distance from the Sun required this forces us to
consider only frequencies well above about 100 GHz (3 mm).

Both the magnification
and the plate scale of a ``Solar Gravitational Telescope'' (SGT) - i.e. the
tangential offset in the image plane corresponding to a given angular
offset in the source - are very large.  Since it is not possible, due to
e.g. propulsion consideration to expect a solar gravitational lens mission to
be able to perform much controlled tangential motion, the observational
program has to be restricted to those that can be accomplished utilizing the
trajectory given by the initial ejection from the inner solar system.  This
will first place very high requirements on the absolute navigation of the
mission and in consequence the trajectory. Second, the source selection for a
SGT is limited to sources that are small (because it takes a long time to
transverse a source), but with interesting smaller-scale (of the order of
the beam) variations which can be explored with one dimensional mapping. 
The source selection further has to be made prior to launch and hence
survey-like observations are out of consideration. A likely observing
scenario is thus to allow the tangential component of the trajectory
velocity to sweep the space craft across the image of the source under
study gathering a one dimensional map as illustrated in Figure
\ref{fig:concept}.  It is questionable whether any source can be found that both,
can be successfully observed with a SGT and, cannot be observed better and cheaper 
using either ground based or Earth orbiting platforms.  

\section{Cruise Science}
\label{sec:cruise}

As with any mission aimed at reaching distances of several hundred AU a SGT
mission can be used to carry out a wide range of \textit{in situ}
measurement while in cruise.  Here we only briefly mention several
possibilities.  For a more detailed description the reader is referred to
earlier studies (e.g. \citet{etc87}):

\paragraph{Heliospheric Science:}  To date only a handful of missions have explored
the outer reaches of the Heliosphere and none has passed beyond it. 
Collecting of \textit{in situ} data on the density, flow and wave structures
of the plasma as well as the magnetic field will help in our understanding
of the solar wind and outer Heliosphere.

\paragraph{The termination of the Heliosphere:}  The zone where the pressures of the
Heliosphere and the local interstellar medium (LISM) come into balance is
of great importance in understanding the interaction between the solar wind
and the LISM as well as of the micro-physics of such interaction regions. 
Low frequency radio observations made by the Voyager 2 space craft now
indicate that this transition region is located at about 85-150 AU in the
upstream direction.  By transversing the termination shock/heliopause/Bow
shock region of the Heliosphere much can be learned about these processes.

\paragraph{LISM science:}  Due to the shielding effect that the Heliosphere has on
the particle fluxes in the ISM, little is known about the detailed density
and composition of the LISM. Reaching outside of the Heliosphere will allow
direct measurements of this region.  What (e.g.) are the
density fluctuations in the interstellar medium?  What are the isotopic
abundances in the LISM? The low energy end of the cosmic ray spectrum is
heavily depleted as seen from Earth and is generally thought to be absorbed
in the Heliosphere.  However, in many processes in the interstellar medium,
such as initiation of chemical reaction networks and heating of dense -
starforming - gas, the exact form of this low energy spectrum is an
important parameter.  This spectrum could be measured once outside the
heliopause.

\paragraph{Outer Solar system bodies:}  If the appropriate telescopes and detectors
are carried on the mission there is a, small, possibility of a chance
encounter (within detection range) of bodies in the Kuiper belt or Oort
cloud.  This would (for Kuiper belt objects) require a trajectory in
the ecliptic plane.  However, as stated above, considerations directly
related to the solar gravitational lensing suggests trajectories out of the
ecliptic plane.

\section{Conclusions}
\label{sec:conclusions}

We have studied a few of the considerations required for a
Solar gravitational lens mission. Specifically we have addressed a number of
problems neglected by earlier studies of such a mission.  The dominant
effects neglected heretofore are those introduced by the plasma in
the solar atmosphere.  This plasma introduces absorption and
scattering (systematic and random) of radiation propagating through it. 
The main effects are to limit the observable wavelength to those well above
about 100 GHz and to increase the effective optical length of the SGT.  We
also note that given realistic constraints on the active control of
the spacecraft trajectory a very limited class of sources can be targeted. 
The targets have to be small enough that interesting changes in their
structure can be expected over several resolution elements ($\sim10^{-7}$as)
while bright enough to be observable very close to the solar disk.  Major
concerns regarding the achievability of the accuracy of the
trajectory required to intercept the optical image of the object under study
($\sim100~{\rm m}$) are raised.

\acknowledgments
We are grateful to Peter Wannier  and Frank Estabrook for the valuable discussions 
during preparation of this manuscript. This work was performed at the Jet
Propulsion Laboratory, California Institute of Technology, under contract with the 
National Aeronautics and Space Administration.





\newpage



\figcaption[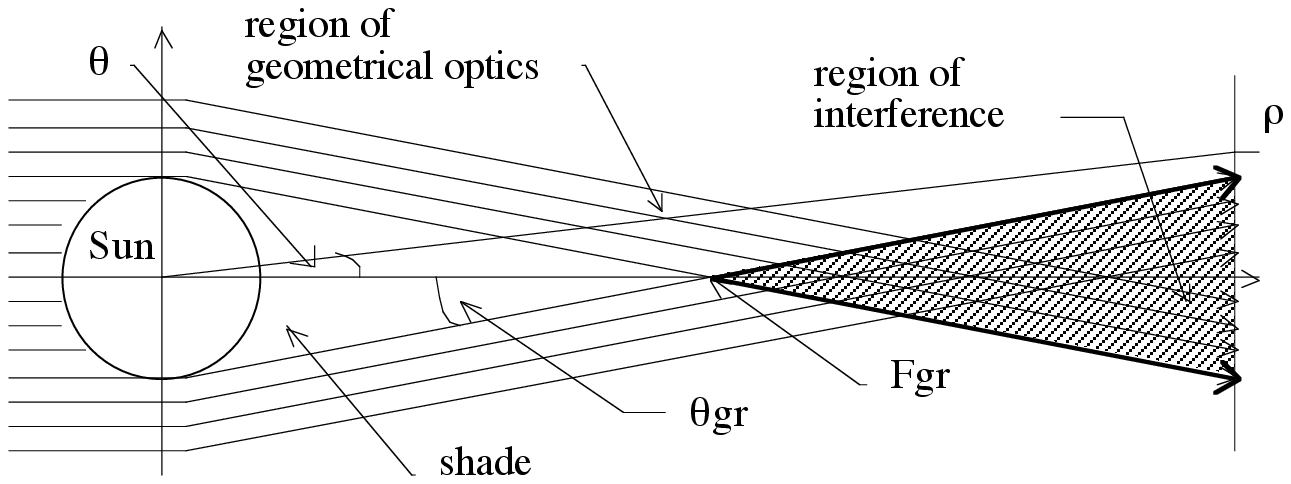]{Different regions of the solar gravity lens.
\label{fig:sl}}

\figcaption[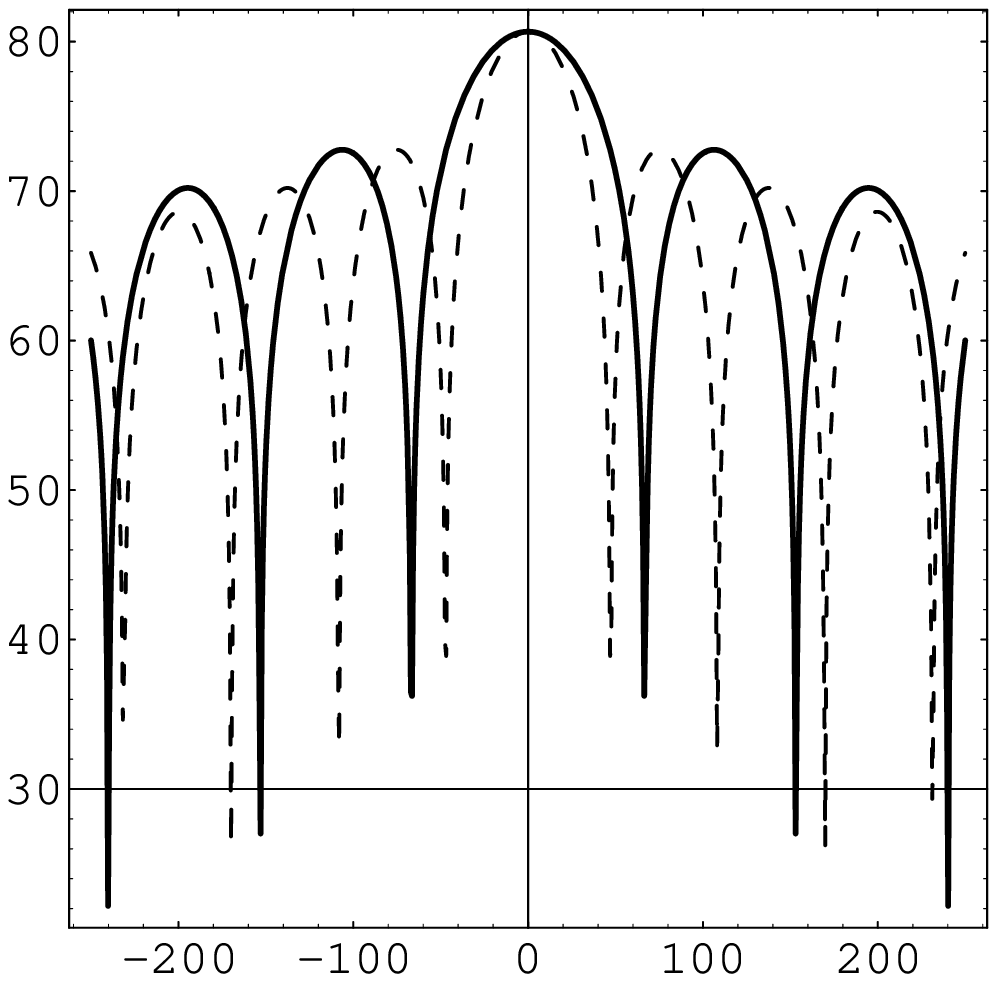]{Gain of the solar gravitational lens as seen in the
image plane as a function of the optical distance ${\cal F}$ and 
possible observational wavelength $\lambda$. The dotted line 
represents gain  for ${\cal F}=600~{\rm AU}$, the thick line  is for 
${\cal F}=1200~{\rm AU}$. \label{fig:gain}}

\figcaption[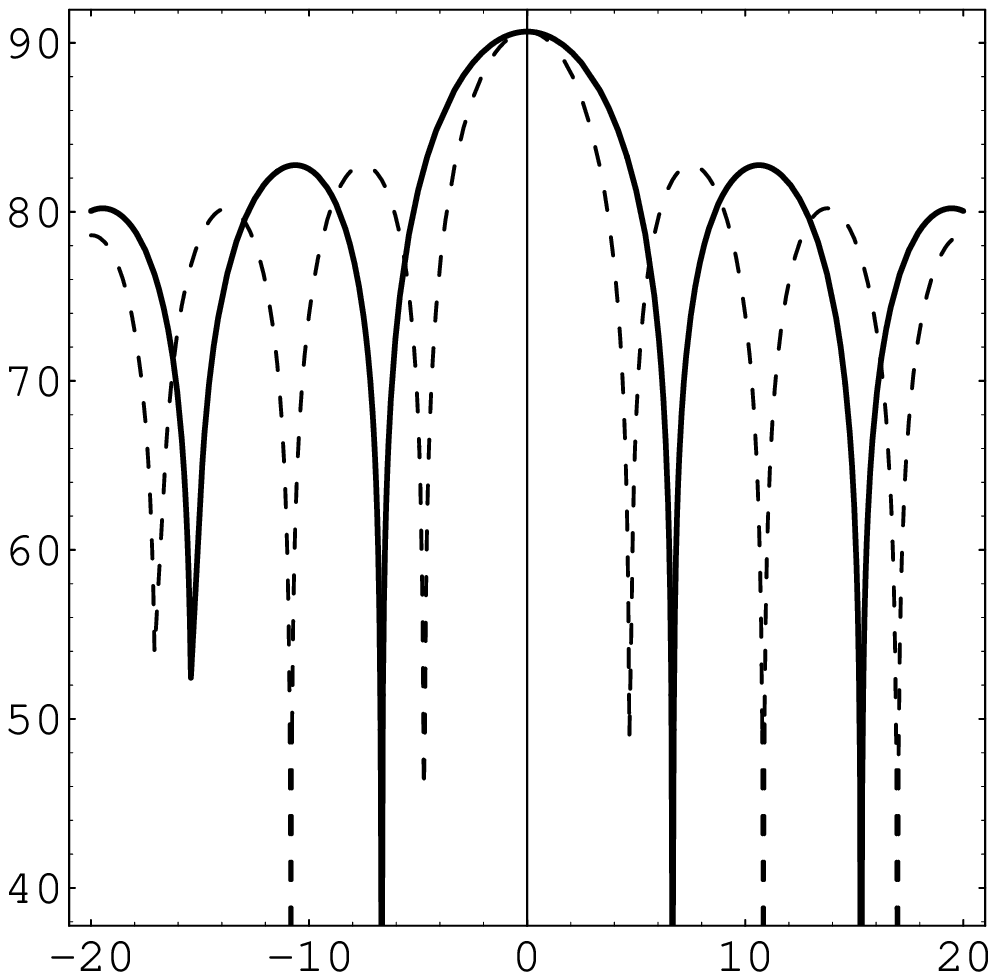]{Refraction of radio-waves in the solar atmosphere.
In this case we adopted the steady-state model, and considered  X
and K-bands radio frequencies (given by lowest dashed and thick lines
correspondingly). The absolute value of the GR bending is also shown 
(dotted line).
      \label{fig:plasma}}

\figcaption[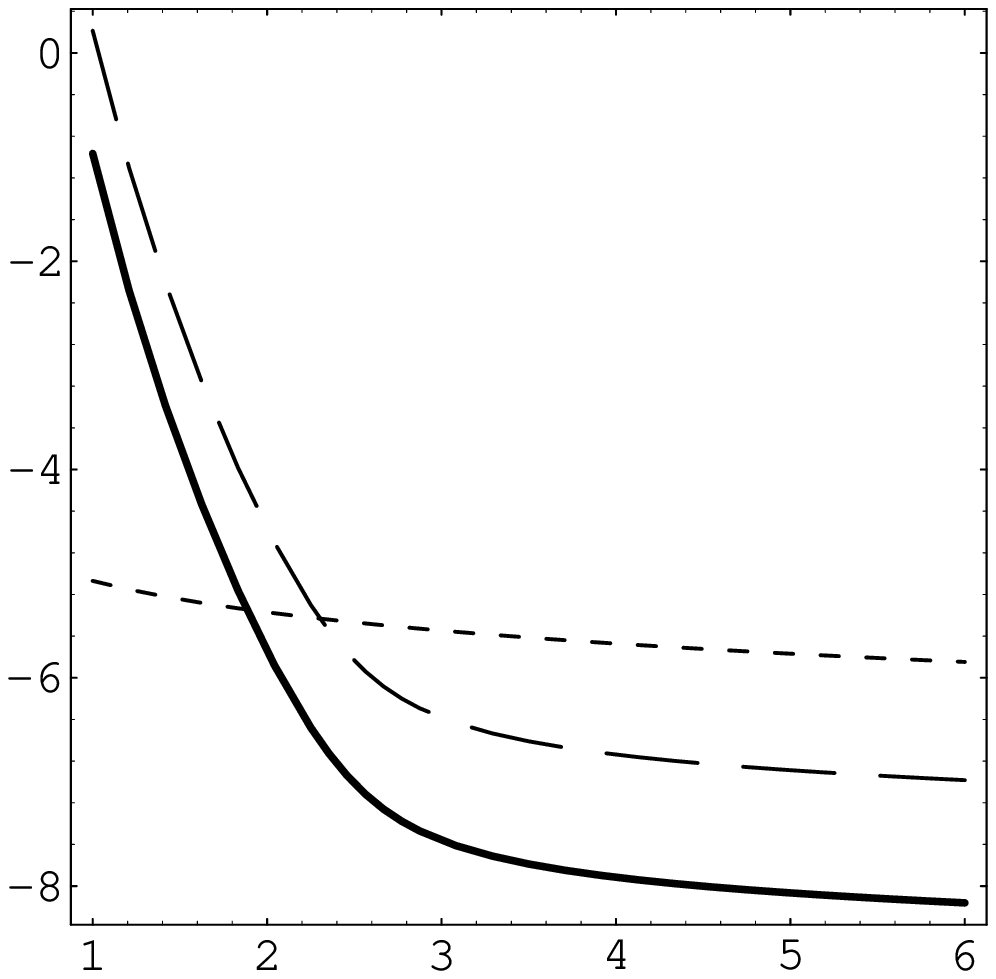]{Effective optical distances for different frequencies 
and impact parameters. The upper curve corresponds to frequency of 170
GHz, then  300 GHz and 500 GHz, and at the  bottom  is for 1 THz. 
      \label{fig:pf}}

\figcaption[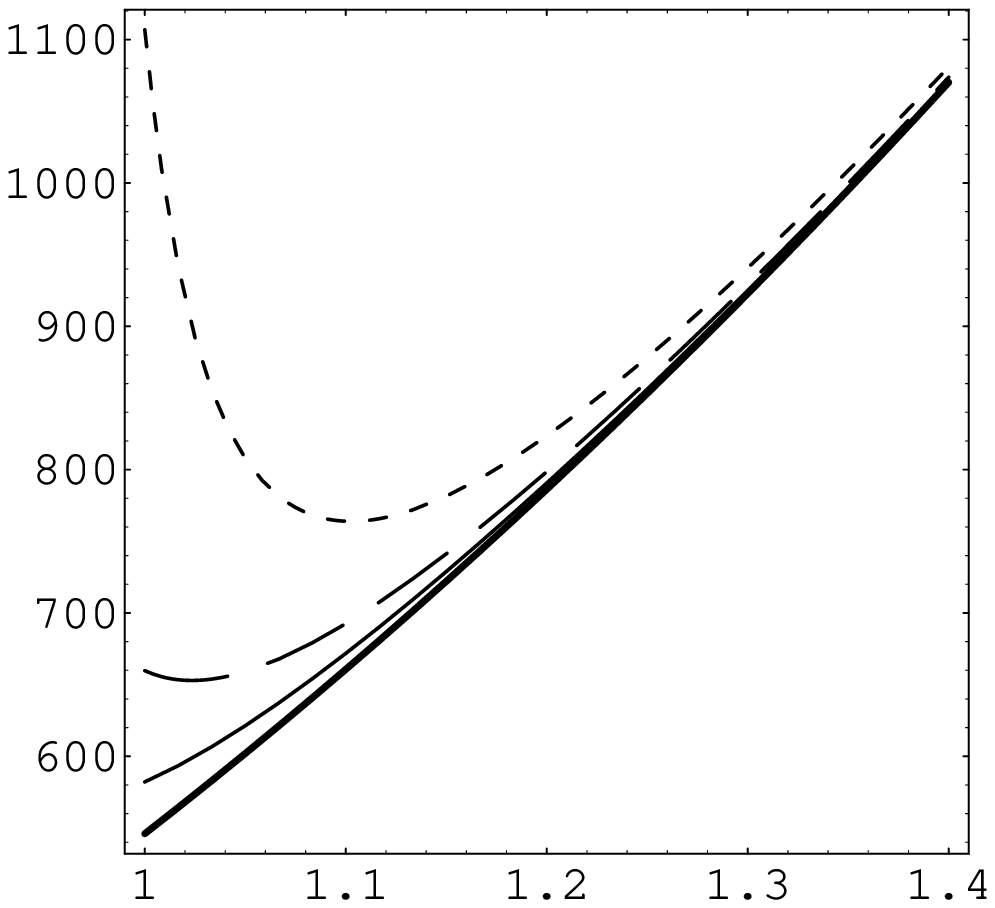]{Conceptual sketch of a possible observing mode.
      \label{fig:concept}}





\clearpage

\begin{table}
\begin{center}
\caption{Estimates of the different terms influencing
the effective optical distance. (Please refer 
to Eq.(\ref{eq:nu}).) \label{tab66}} \vskip 10pt
{\small
\begin{tabular}
{|c|c|c|c|c|c|c|} \hline\hline
$ {b}/{{\cal R}_\odot }$  &${\cal F}_{\tt gr}$, AU & $\sim
\Big({{\cal R}_\odot / b}\Big)^{15}$   & $\sim\Big({{\cal R}_\odot /
b}\Big)^5$ &$\sim\Big({{\cal R}_\odot /
b}\Big)$  \\\hline
1     & 546  & 2952 & 228 & 1.1  \\
1.05  & 601  & 1426 & 178 & 1    \\
1.1   & 660  &  704 & 141 & 1    \\
1.15  & 722  &  364 & 113 & 0.96 \\
1.2   & 785  &  192 &  92 & 0.92 \\
1.25  & 853  &  104 &  75 & 0.88 \\
1.3   & 922  &   58 &  62 & 0.85 \\
1.35  &  996    &   32 &  51 & 0.81 \\
1.4   & 1070 &   19 &  43 & 0.78 \\
1.5   & 1228 &    7 &  30 & 0.73 \\
1.6   & 1397 &    2 &  22 & 0.69 \\\hline
\end{tabular}
}
\end{center}
\end{table}

\clearpage

\begin{figure}[ht]
 \begin{center}
    \epsfig{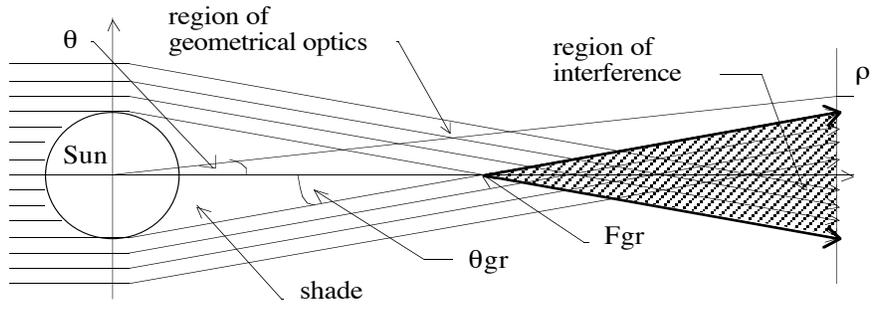}
     \caption{{\small Different regions of the solar gravity lens.}
      \label{fig:sl}}
 \end{center}
\end{figure}

\clearpage

\vskip 10pt
\begin{figure}[ht]
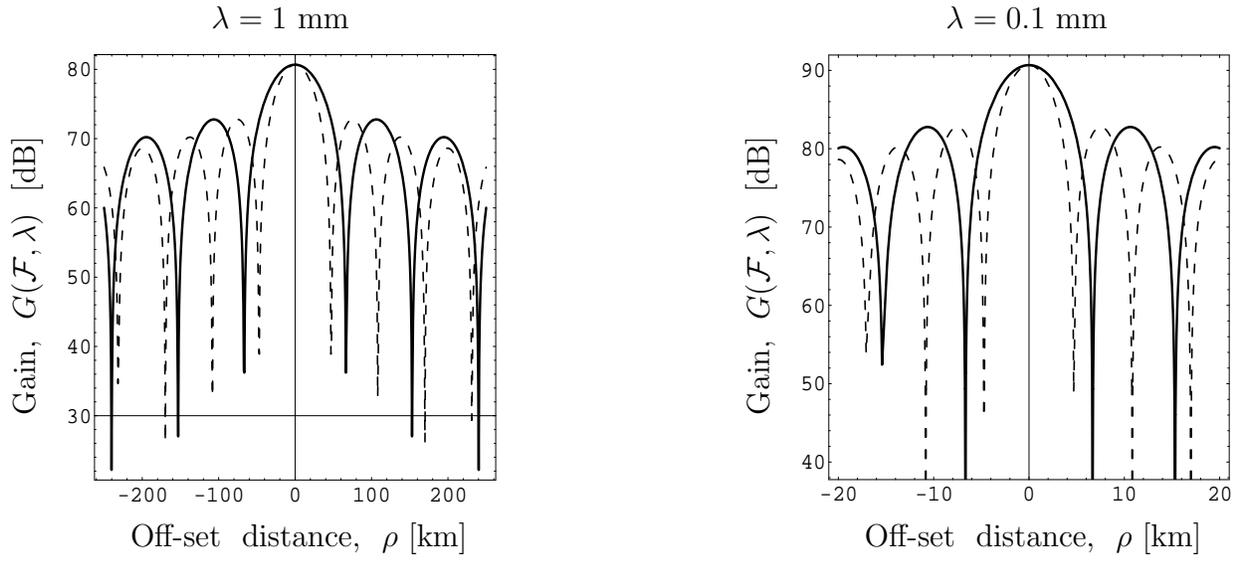

 \begin{center}
\rotatebox{90}{\hskip 50pt  Gain,   ~$G({\cal F},\lambda)$ ~[dB]}
\hskip -40pt
\begin{minipage}[b]{.46\linewidth}
\rotatebox{0}{\hskip 100pt  $\lambda=1~{\rm mm}$}
\vskip -20pt
\centering \epsfig{figure=fg2.eps,width=70mm,height=70mm}
\rotatebox{0}{\hskip 50pt  Off-set ~distance, ~$\rho$~[km]}
\end{minipage}
\hfill
\rotatebox{90}{\hskip 50pt  Gain,   ~$G({\cal F},\lambda)$ ~[dB]}
\hskip -40pt
\begin{minipage}[b]{.46\linewidth}
\rotatebox{0}{\hskip 100pt  $\lambda=0.1~{\rm mm}$}
\vskip -20pt
\centering \epsfig{figure=fg3.eps,width=70mm,height=70mm}
\rotatebox{0}{\hskip 50pt  Off-set ~distance, ~$\rho$~[km]}
\end{minipage}
     \caption{{\small Gain of the solar gravitational lens as seen in the
image plane as a function of the optical distance ${\cal F}$ and 
possible observational wavelength $\lambda$. The dotted line represents
gain  for ${\cal F}=600~{\rm AU}$, the thick line  is for ${\cal
F}=1200~{\rm AU}$.}
      \label{fig:gain}}
 \end{center}
\end{figure}

\clearpage

\begin{figure}[ht]
 \begin{center}
\rotatebox{90}{\hskip 30pt  Refraction ~angle,  
~${\rm log}_{10}\theta_{\tt pl}(b,\nu)$ ~[rad]}
\hskip -50pt
    \epsfig{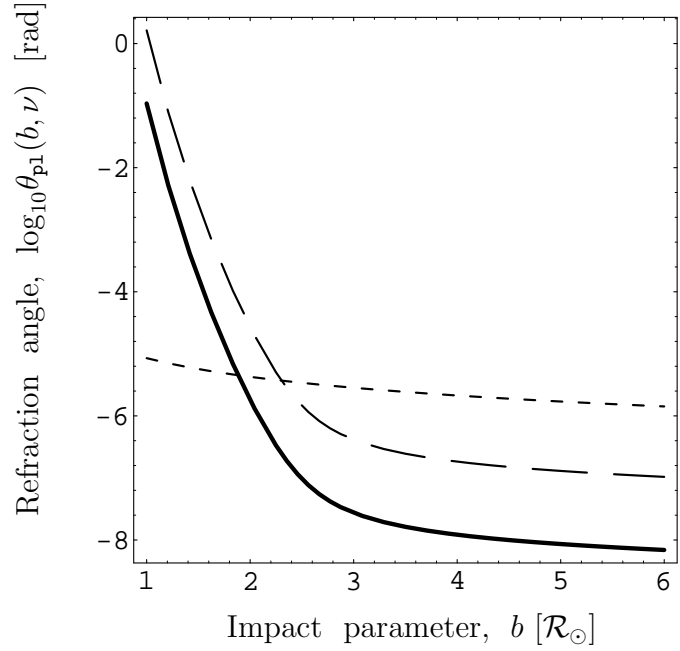}

\rotatebox{0}{\hskip 50pt Impact ~parameter, ~$b~[{\cal R}_\odot]$}
     \caption{{\small Refraction of radio-waves in the solar atmosphere.
In this case we adopted the steady-state model, and considered  X
and K-bands radio frequencies (given by lowest dashed and thick lines
correspondingly). The absolute value of the GR bending is also shown 
(dotted line).}
      \label{fig:plasma}}
 \end{center}
\end{figure}

\clearpage

\begin{figure}[ht]
 \begin{center}
\rotatebox{90}{\hskip 20pt Effective ~distance, 
~${\cal F}_{\tt gr+pl}(b,\nu)~[{\rm AU}]$}
\hskip -45pt
    \epsfig{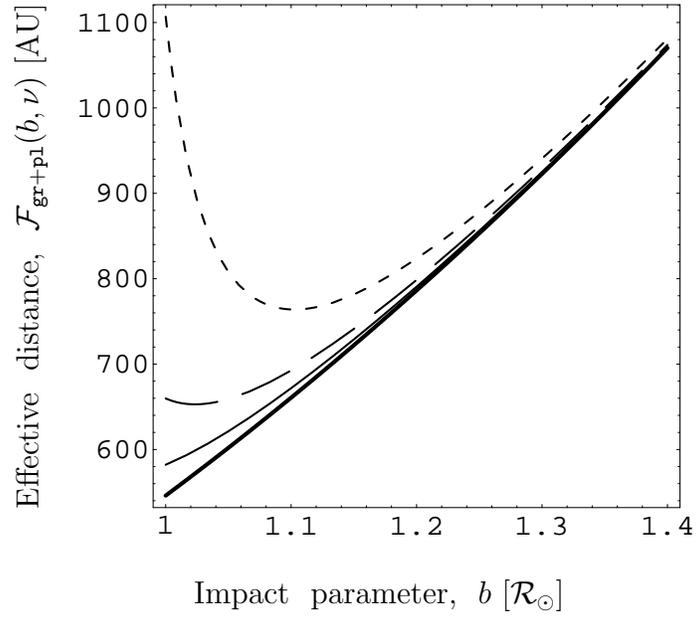}

\rotatebox{0}{\hskip 20pt Impact ~parameter, ~$b~[{\cal R}_\odot]$}
    \caption{{\small Effective optical distances for different frequencies 
and impact parameters. The upper curve corresponds to frequency of 170
GHz, then  300 GHz and 500 GHz, and at the  bottom  is for 1 THz.  }
      \label{fig:pf}}
 \end{center}
\end{figure}

\clearpage

\clearpage

\begin{figure}[ht]
 \begin{center}
    \epsfig{figure=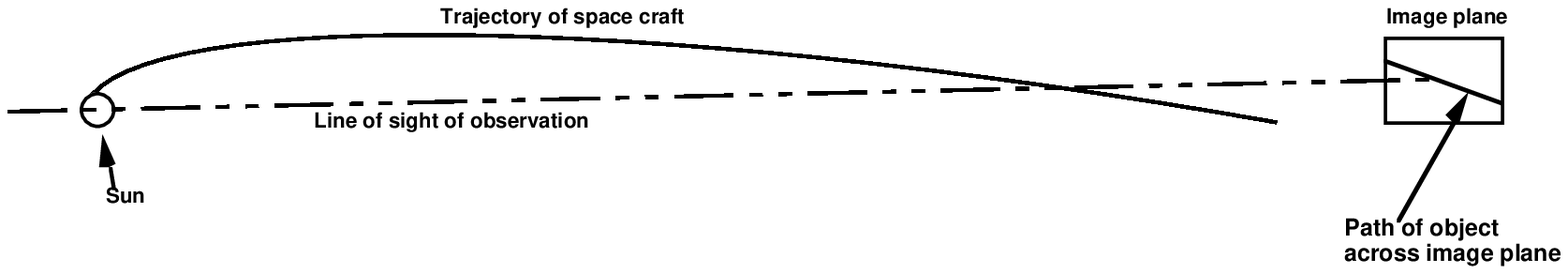,width=145mm,height=40mm}
     \caption{{\small Conceptual sketch of a possible observing mode.}
      \label{fig:concept}}
 \end{center}
\end{figure}

\end{document}